\documentstyle[aps]{revtex}

\begin{document}
\draft
\title{OPTICALLY DETECTED MAGNETOPHONON RESONANCES IN POLAR SEMICONDUCTORS}
\author{G.-Q. Hai}
\address{Instituto de F\'\i sica de S\~ao Carlos, Universidade de S\~ao Paulo, \\
13560-970, S\~ao Carlos, SP, Brazil}
\author{F. M. Peeters}
\address{Departement Natuurkunde, Universiteit Antwerpen (UIA),\\
B-2610 Antwerpen, Belgium}
\maketitle
\date{\today}

\begin{abstract}
Magnetophonon resonances are found for $\omega _{c}=\omega _{LO}/N$ with $N=1,2,3$ ....
in the polaron cyclotron resonance (CR)
linewidth and effective mass of bulk polar semiconductors.
The CR mass and the linewidth are obtained from 
the full polaron magneto-optical absorption spectrum which are calculated 
using the memory function technique.
The amplitude of the resonant peak in the linewidth can
be described by an exponential law at low temperature.
\end{abstract}


\section{INTRODUCTION}

The magnetophonon resonance (MPR) effect occurs when two Landau levels are a
phonon energy apart which leads to a resonant scattering due to emission or
absorption of phonons. Since the pioneering work by Gurevich and Firsov [1],
this effect has been extensively studied in bulk [2-4] as well as
low-dimensional semiconductor systems.[5-10] The resonant character makes it
a powerful spectroscopic tool. Magnetophonon resonances have been used to
obtain information on band structure parameters, such as the effective mass
and the energy levels, and on the electron-phonon interaction. 
The vast majority of work on MPR was done on the
transport properties of semiconductors, usually the magnetoresistance, which 
inevitably involves a complicated averaging of scattering processes. The
oscillations in the magnetoresistance are the results of a combination of
scattering and broadening processes which can lead to a quite complicated
dependence of the resonance amplitudes on doping, sample structure, carrier
concentration, and temperature. However, the MPR can also be observed
directly through a study of the electron cyclotron resonance (CR) linewidth
and effective mass, i.e., the so called optically detected MPR (ODMPR).
This was demonstrated by Barnes {\it et al.} [11] for a 
two-dimensional (2D) semiconductor system formed in GaAs/AlGaAs
heterojunctions.
The ODMPR allows one to make
quantitative measurements of the scattering strength for specific Landau
levels and yields direct information on the nature of the electron-phonon
interaction in semiconductors.

In this work, we extend the theory for ODMPR to three-dimensional (3D)
systems and present the first theoretical study of the magnetophonon
resonances in the frequency-dependent conductivity in bulk polar
semiconductors. Our calculations show strong oscillations of both the
linewidth and the effective mass in a 3D system of GaAs which indicate that
the ODMPR should also be observed experimentally in bulk polar
semiconductors. 

The present paper is organized as follows. In Sec. II, \ we
present our theoretical formulation of the problem. The numerical results
and discussions are given in Sec. III, and we summerize our results in Sec.
IV.

\section{Theoretical Framework}

Magnetophonon resonance is essentially a single-particle effect and,
consequently, can be treated as a one-polaron problem. We consider a polar
semiconductor in a uniform magnetic field {\bf B} directed along the $z$%
-axis. The system under consideration can be described by the following
Hamiltonian, 
\begin{equation}
H=H_{e}+H_{ph}+H_{int}
\end{equation}
with 
\begin{equation}
H_{e}={\frac{1}{2m_{b}}}(\vec{p}+e\vec{A})^{2},
\end{equation}
and 
\begin{equation}
H_{ph}=\sum_{\vec{q}}\hbar \omega _{\vec{q}}(a_{\vec{q}}^{\dagger }a_{\vec{q}%
}+{{\frac{_{1}}{{^{2}}}}})\ ,
\end{equation}
where $m_{b}$ is the bare electron effective mass, the vector potential $%
\vec{A}=B/2(-y,x,0)$ is choosen in the symmetrical Coulomb gauge, $\vec{q}$~
($\vec{r}$) the momentum (position) operator of the electron, $a_{\vec{q}%
}^{\dagger }$ $(a_{\vec{q}})$ the creation (annihilation) operator of an
optical-phonon with wave vector $\vec{q}$ and energy $\hbar \omega _{\vec{q}%
} $. The electron-phonon interaction Hamiltonian $H_{int}$ is given by the
Fr\"{o}hlich interaction Hamiltonian 
\begin{equation}
H_{int}=\sum_{\vec{q}}(V_{\vec{q}}a_{\vec{q}}e^{i\vec{q}\cdot \vec{r}}+V_{%
\vec{q}}^{\ast }a_{\vec{q}}^{\dagger }e^{-i\vec{q}\cdot \vec{r}})\ ,
\end{equation}
where 
\begin{equation}
V_{\vec{q}}=-i\hbar \omega _{LO}\left( {\frac{\hbar }{2m_{b}\omega _{LO}}}%
\right) ^{1/4}\sqrt{\frac{4\pi \alpha }{Vq^{2}}}\;,
\end{equation}
and $\alpha $ is the electron-LO-phonon coupling constant.

First, we calculate the optical aborption spectrum of
the polaron in the presence of a magnetic field
from which we are able to investigate the
polaron CR spectrum and the corresponding MPR effects.
For convenience we use units such that $%
\hbar =m_{b}=\omega _{LO}=1.$ Within linear response theory the
frequency-dependent magneto-optical absorption spectrum for cyclotron
resonance [12-14] is given by 
\begin{equation}
A(\omega )=-{\frac{1}{2}}{\frac{{\rm Im}\Sigma (z)}{[\omega -\omega _{c}-%
{\rm Re}\Sigma (z)]^{2}+[{\rm Im}\Sigma (z)]^{2}}},  \label{absorption}
\end{equation}
where $\omega _{c}=eB/m_{b}$ is the unperturbed electron cyclotron
frequency, $\Sigma (z)$ is the so called memory function, and $z=\omega
+i\gamma $ where $\gamma $ is a broadening parameter. Notice
that $\gamma $ is introduced semi-empirically to remove the divergence of
the Landau level density of states. 
We take $\gamma$ as a constant. For the
magneto-optical absorption spectrum in the Faraday (active-mode)
configuration, which corresponds to the cyclotron resonance experiments, the
memory function is given by [13] 
\begin{equation}
\Sigma (z)={\frac{1}{2m_{b}}}\sum_{\vec{q}}q^{2}|V_{\vec{q}}|^{2}F_{\vec{q}%
}(z),
\end{equation}
with 
\begin{equation}
F_{\vec{q}}(z)=-{\frac{2}{z}}\int_{0}^{\infty }dt(1-e^{izt}){\rm Im}\langle
\lbrack b_{\vec{q}}(t),b_{\vec{q}}^{\dagger }(0)]\rangle ,
\end{equation}
where $b_{\vec{q}}=a_{\vec{q}}e^{i{\vec{q}}\cdot {\vec{r}}}$, and the
correlation function is given by 
\begin{equation}
\langle \lbrack b_{\vec{q}}(t),b_{\vec{q}}^{\dagger }(0)]\rangle
=[1+n(\omega _{LO})]e^{-i\omega _{LO}t}S^{\ast }(-\vec{q},t)-n(\omega
_{LO})e^{-i\omega _{LO}t}S(\vec{q},t),
\end{equation}
where 
\begin{equation}
n(\omega _{LO})={\frac{1}{e^{\beta \omega _{LO}}-1}} , 
\end{equation}
is the number of LO-phonons and 
\begin{equation}
S(\vec{q},t)=\langle e^{i\vec{q}\cdot \vec{r}(t)}e^{-i\vec{q}\cdot \vec{r}%
(0)}\rangle ,
\end{equation}
is the space Fourier transform of the electron density-density correlation
function. In Eq. (10) $\beta =1/k_{B}T$, where $k_{B}$ is the Boltzmann
constant. For a weak electron-LO-phonon coupling system, i.e. $\alpha <<1$,
the density-density correlation function is calculated for a free electron in a
magnetic field which is given by 
\begin{equation}
S(\vec{q},t)=e^{q_{z}^{2}D(t)}e^{-q^{2}D_{H}(t)},
\end{equation}
where 
\begin{equation}
D(t)=\frac{1}{2}(-it+t^{2}/\beta ) ,
\end{equation}
and 
\[
D_{H}(t)={\frac{1}{2\omega _{c}}}[1-e^{i\omega _{c}t}+4n(\omega _{c})\sin
^{2}(\omega _{c}t/2)] ,
\]

From the above equations, we obtain the memory function for $\gamma =0$. 
The calculation proceeds along the lines of a similar calculation which was
presented in Ref. [13]. The results for the memory function are
\begin{eqnarray}
{\rm Re}\Sigma (\omega ) &=&\frac{\alpha \sqrt{\beta }}{2\pi \omega }{\frac{%
\omega _{c}\tanh (\beta \omega _{c}/2)}{\sinh (\beta /2)}}\sum_{n,n^{\prime
}=0}^{\infty }{\frac{\left[ 2\cosh (\beta \omega _{c}/2)\right]
^{-(n+n^{\prime })}}{n!n^{\prime }!}}\int_{0}^{\infty }{\frac{dx}{x}}%
E_{n+n^{\prime }+1}\left( {\frac{x^{2}}{\omega _{c}\tanh (\beta \omega
_{c}/2)}}\right)  \nonumber \\
&&\times \{\exp \left( {\frac{\beta \omega _{nn^{\prime }}}{2}}\right) \{2{%
D\left( {\frac{\sqrt{\beta }x}{2}}+{\frac{\sqrt{\beta }}{2x}}\omega
_{nn^{\prime }}\right) }  \nonumber \\
{} &{}&-{D\left( {\frac{\sqrt{\beta }x}{2}}+{\frac{\sqrt{\beta }}{2x}}%
(\omega _{nn^{\prime }}+\omega )\right) }-{D\left( {\frac{\sqrt{\beta }x}{2}}%
+{\frac{\sqrt{\beta }}{2x}}(\omega _{n^{\prime }n}-\omega )\right) }\} 
\nonumber \\
&&+\exp \left( -{\frac{\beta \omega _{n^{\prime }n}}{2}}\right) \{2{D\left( {%
\frac{\sqrt{\beta }x}{2}-\frac{\sqrt{\beta }}{2x}\omega _{n^{\prime }n}}%
\right) }  \nonumber \\
&&{-D\left( {\frac{\sqrt{\beta }x}{2}-\frac{\sqrt{\beta }}{2x}}(\omega
_{n^{\prime }n}-\omega )\right) }-{D\left( {\frac{\sqrt{\beta }x}{2}-\frac{%
\sqrt{\beta }}{2x}}(\omega _{n^{\prime }n}+\omega )\right) }\}\},
\end{eqnarray}
and 
\begin{eqnarray}
{\rm Im}\Sigma (\omega )=- &&\frac{\alpha \sqrt{\beta }}{4\sqrt{\pi }\omega }%
{\frac{\omega _{c}\sinh (\beta \omega /2)\tanh (\beta \omega _{c}/2)}{\sinh
(\beta /2)}}\sum_{n,n^{\prime }=0}^{\infty }{\frac{\left[ 2\cosh (\beta
\omega _{c}/2)\right] ^{-(n+n^{\prime })}}{n!n^{\prime }!}}  \nonumber \\
&&\times \int_{0}^{\infty }{\frac{dx}{x}}E_{n+n^{\prime }+1}\left( {\frac{%
x^{2}}{\omega _{c}\tanh (\beta \omega _{c}/2)}}\right)  \nonumber \\
&&\times \left[ \exp \left( -{\frac{\beta x}{4}}-{\frac{\beta }{4x}}(\omega
_{nn^{\prime }}-\omega )\right) +\exp \left( -{\frac{\beta x}{4}}+{\frac{%
\beta }{4x}}(\omega _{n^{\prime }n}-\omega )\right) \right] ,
\end{eqnarray}
where, $\omega _{nn^{\prime }}=1+(n-n^{\prime })\omega _{c}$, $D(x)$ is the
Dawson integral function, and 
\begin{equation}
E_{n}=\int_{0}^{\infty }dt\;{\frac{t^{n}e^{-t}}{t+x}}.
\end{equation}

In the case of non zero broadening, i.e. $\gamma \neq 0,$ 
the calculation is more tedious. We
obtain the following results for the memory function, 
\begin{equation}
{\rm Re}\Sigma (\omega )=-{\frac{\alpha }{\sqrt{2}\pi (\omega ^{2}+\gamma
^{2})}}\left[ \omega I_{1}(\omega )+\gamma I_{2}(\omega )\right] ,
\end{equation}
and 
\begin{equation}
{\rm Im}\Sigma (\omega )={\frac{\alpha }{\sqrt{2}\pi (\omega ^{2}+\gamma
^{2})}}\left[ \omega I_{2}(\omega )+\gamma I_{1}(\omega )\right] ,
\end{equation}
with 
\begin{eqnarray}
I_{1}(\omega ) &=&-\sqrt{2\beta }{\frac{\omega _{c}\tanh (\beta \omega
_{c}/2)}{\sinh (\beta /2)}}\sum_{n,n^{\prime }=0}^{\infty }{\frac{\left[
2\cosh (\beta \omega _{c}/2)\right] ^{-(n+n^{\prime })}}{n!n^{\prime }!}}%
\int_{0}^{\infty }{\frac{dx}{x}}E_{n+n^{\prime }+1}\left( {\frac{x^{2}}{%
\omega _{c}\tanh (\beta \omega _{c}/2)}}\right)  \nonumber \\
&&\times \{\exp \left( {\frac{\beta \omega _{nn^{\prime }}}{2}}\right) \{{%
D\left( {\frac{\sqrt{\beta }x}{2}}+{\frac{\sqrt{\beta }}{2x}}\omega
_{nn^{\prime }}\right) }  \nonumber \\
{} &{}&{-{\frac{\sqrt{\pi }}{4}}}\left[ {{\rm Im}W\left( {\frac{\sqrt{\beta }%
x}{2}}+{\frac{\sqrt{\beta }}{2x}}(\omega _{nn^{\prime }}+\omega +i\gamma
)\right) }+{\rm Im}W{\left( {\frac{\sqrt{\beta }x}{2}}+{\frac{\sqrt{\beta }}{%
2x}}(\omega _{n^{\prime }n}-\omega +i\gamma )\right) }\right] \}  \nonumber
\\
&&+\exp \left( -{\frac{\beta \omega _{n^{\prime }n}}{2}}\right) \{{D\left( {%
\frac{\sqrt{\beta }x}{2}-\frac{\sqrt{\beta }}{2x}\omega _{n^{\prime }n}}%
\right) }  \nonumber \\
&&{-{\frac{\sqrt{\pi }}{4}}}\left[ {{\rm Im}W\left( {\frac{\sqrt{\beta }x}{2}%
-\frac{\sqrt{\beta }}{2x}}(\omega _{n^{\prime }n}-\omega -i\gamma )\right) }+%
{\rm Im}W{\left( {\frac{\sqrt{\beta }x}{2}-\frac{\sqrt{\beta }}{2x}}(\omega
_{n^{\prime }n}+\omega -i\gamma )\right) }\right] \}\},
\end{eqnarray}
and 
\begin{eqnarray}
I_{2}(\omega )=-\frac{\sqrt{\pi \beta }}{2\sqrt{2}}{\frac{\omega _{c}\tanh
(\beta \omega _{c}/2)}{\sinh (\beta /2)}} &&\sum_{n,n^{\prime }=0}^{\infty }{%
\frac{\left[ 2\cosh (\beta \omega _{c}/2)\right] ^{-(n+n^{\prime })}}{%
n!n^{\prime }!}}\int_{0}^{\infty }{\frac{dx}{x}}E_{n+n^{\prime }+1}\left( {%
\frac{x^{2}}{\omega _{c}\tanh (\beta \omega _{c}/2)}}\right)  \nonumber \\
\times \{\exp \left( {\frac{\beta \omega _{nn^{\prime }}}{2}}\right) &&[{\rm %
Re}W\left( {\frac{\sqrt{\beta }x}{2}}+{\frac{\sqrt{\beta }}{2x}}(\omega
_{nn^{\prime }}+\omega +i\gamma )\right)  \nonumber \\
&&-{\rm Re}W\left( {\frac{\sqrt{\beta }x}{2}}+{\frac{\sqrt{\beta }}{2x}}%
(\omega _{nn^{\prime }}-\omega +i\gamma )\right) ]  \nonumber \\
+\exp \left( -{\frac{\beta \omega _{n^{\prime }n}}{2}}\right) &&[{\rm Re}%
W\left( {\frac{\sqrt{\beta }x}{2}}-{\frac{\sqrt{\beta }}{2x}}(\omega
_{n^{\prime }n}-\omega -i\gamma )\right)  \nonumber \\
&&-{\rm Re}W\left( {\frac{\sqrt{\beta }x}{2}}-{\frac{\sqrt{\beta }}{2x}}%
(\omega _{n^{\prime }n}+\omega -i\gamma )\right) ]\},
\end{eqnarray}
where $W(z)=e^{-z^{2}}{\rm erfc}(-iz)$ is the complex error function.

\section{numerical results and discussions}

In this section, we present our numerical results on the
magneto-optical aborption spectra and study the magnetophonon resonant
effects. As an example, we apply
our theory to the semiconductor GaAs where $\alpha =0.07$. First, we show 
numerical results for $T=77$ K and level broadening
parameter $\gamma =0$. Due to the importance of the memory function in the
absorption spectrum, we plot the real and imaginary parts of the memory
function in Fig. 1(a) and 1(b), respectively, as a function of frequency at
three different magnetic fields. We see that, at $\omega =|\omega _{LO}-n\omega
_{c}|$ ($n=0,1,2,$...), Re$\Sigma (\omega )$ exhibits a jump while Im$\Sigma
(\omega )$ diverges logarithmically. The 
discontinuity of Re$\Sigma (\omega )$ and the divergency in Im$\Sigma (\omega)$
reflects the resonant coupling between the
state $E_{0}+\omega _{LO}$ and Landau level $E_{n}=(1/2+n)\omega _{c}.$ The
stronger this coupling, the larger the discontinuity in Re$\Sigma (\omega )$.
Actually, the real part of the memory function Re$\Sigma (\omega )$ is
responsible for the shift in the observed CR energy which is due to the
electron-phonon interaction. While, the imaginary part leads to a broadening of
the spectrum which is a result of scattering. When Im$\Sigma (\omega )=0$
like in a 2D system, the
absorption is a $\delta $ function and its position is determined by the
equation $\omega _{c}^{\ast }-\omega _{c}-{\rm Re}\Sigma (\omega _{c}^{\ast
})=0$. Fig. 1(b) shows that for a bulk system Im$\Sigma (\omega )$
is non zero which reflects the 3D character of the electron states. The
scattering which is mainly in the direction parallel to the magnetic field results in
a finite Im$\Sigma (\omega )$ and consequently a finite linewidth even for $\gamma
=0$. In Fig. 1(c), we show the corresponding magneto-optical absorption
spectra. The position of the aborption peaks corresponds to the cyclotron
resonance frequency $\omega _{c}^{\ast }$ at which the cyclotron resonance
occurs. We notice an asymmetric double peak structure around $\omega =\omega
_{LO}/2$ for $\omega _{c}=\omega _{LO}/2$ (the solid curve) and the
absorption becomes zero at $\omega =\omega _{c}=\omega _{LO}/2.$
The zeros in the absorption spectrum are a consequence of the divergences in 
Im$\Sigma (\omega)$ and which can be traced back to the divergent nature of the
density of states. The double peak structure is a consequence of
the magnetophonon resonance which leads to an anticrossing 
behavior in the CR spectrum. When the CR frequency $\omega _{c}$ deviates
from $\omega _{LO}/N,$ this splitting becomes very weak and difficult to 
observe in the aborption spectrum. As we will see below, however, the
magnetophonon resonance will strongly affect the linewidth of the
magneto-optical aborption and the CR mass. From the dash and the dotted
curves, we notice that the absorption peak appears at 
a frequency $\omega _{c}^{\ast
}<\omega _{c}$ due to the polaron effect which shifts the
cyclotron frequency to lower frequencies. The latter is often interpreted as an
increase of the cyclotron mass, i.e. $\omega _{c}^{\ast} = eB/m^*c$.
In Fig. 2, we show the
absorption spectrum around (a) $\omega_c =\omega _{LO}/2$ and (b) $\omega_c
=\omega _{LO}/3$. The double peak structure disappears when $\omega_c $
deviates from $\omega _{LO}/N$. The aborption spectra also demonstrate 
clearly a nonlinear magnetic field dependence of the peak position and linewidth
around $\omega _{c}/N.$ \ 

Fig. 3 demonstrates the effect of the broadening parameter $%
\gamma $ on the absorption spectrum. Notice that, with increasing $\gamma $:
i) the double peak structure disappears for $\gamma >0.01\omega _{LO}$,
ii) the zero in the absorption spectrum disappears when $\gamma > 0$, and 
iii) the position
of the absorption peak shifts to higher frequency. This indicates that the
anticrossing behavior in the CR spectrum will be difficult to be observed 
experimently at $\omega _{c}=\omega _{LO}/2$ due to broadening effects which
are a consequence of scattering on e.g. impurities and acoustical phonons.

As soon as the polaron CR frequency $\omega _{c}^{\ast }$ is determined from 
the postion of the magneto-optical absorption peak, the CR mass of the
polaron is obtained by 
\begin{equation}
m^{\ast }/m_{b}=\omega _{c}/\omega _{c}^{\ast }\;.
\end{equation}
The numerical results of the polaron CR mass and the FWHM (full width at
half maximum) for $\gamma =0.05\omega _{LO}$ are plotted as a function of
the unperturbed CR frequency at different temperatures in Figs. 4(a) and
4(b), respectively. One observes that the polaron CR mass is an oscillating 
function of the magnetic field. Fig. 4(b) shows that the FWHM of the polaron
magneto-optical absorpion spectrum reaches a local maximum at $\omega
_{c}=\omega _{LO}/N$ where the polaron mass has an inflection point.
This result demonstrates the derivative-like relation
between the polaron CR mass and the linewidth which are due to the fact that
the real and imaginary part of the memory function are related to each other
through a Kramers-Kronig relation. One finds
that, for temperature $T<100$ K, the resonance grows rapidly with increasing 
$T$. This effect can lead to a direct measurement of the optical-phonon
scattering rate. We notice also an overall increase of the linewidth with
temperature, but an overall decrease of the effective mass when $T>80$ K.
The resonant position is slightly larger than the non-interacting 
resonant condition $%
\omega _{c}=\omega _{LO}/N$ and is almost independent of temperature. A
detailed analysis indicates that, at $N=2$ and $3,$ the peak position both in
the FWHM and in the derivative of the CR mass is at $0.504\omega _{LO}$
and $0.336\omega _{LO}$, respectively. Experimentally this position
determines the so called fundamental field $B_{0}=m^{\ast }\omega _{LO}/e$
which is an important quantity to study the effective mass,
nonparabolicity of the energy band, as well as the LO-phonon frequency.
The linewidth is a direct measure of the lifetime of the state. Notice that the
conventional MPR occurs in the resistivity, which is given by $\rho _{zz}=-$ Im$%
\Sigma (\omega =0)$. But ODMPR is related to both the real and imaginary
part of the memory function which occurs for $\omega \ne 0$ and 
is a dynamical MPR. 

Fig. 5 shows the CR mass oscillation amplitude at $\omega _{c}/\omega
_{LO}=1/2$ and $1/3$ as a function of temperature. With increasing
temperature, the number of phonons increases and consequently, the oscillation
amplitude increases. On the other hand, the background
electron-phonon scattering (coupling) increases which results in a
suppression of the oscillation amplitude. Fig. 6 shows an activation plot
of the amplitude of the resonant peak in the FWHM at $\omega _{c}/\omega
_{LO}=1/2$ and $1/3$ as a function of $T^{-1}$. We find that, for the
resonance around $N=2$, the linewidth can be described rather well by the
exponential law $\exp (-\hbar \omega _{LO}/2kT)$ for $T<240$ K, while that
around $N=3$ can be described by $\exp (-2\hbar \omega _{LO}/3kT)$ for $%
T<140 $ K.
This exponential behavior can be understood as follows. MPR is proportional to
the number of LO-phonons which are present and therefore should increase as
$n(\omega_{LO})$. On the other hand, thermal broadening of the Landau levels,
which is proportional to $n(\omega_c)$, will diminish the resonant structure
in $\Delta FWHM$. Thus this contribution decreases the resonant character and
consequently we expect that $\Delta FWHM \sim n(\omega_{LO})/n(\omega_c)
\approx exp(-\hbar(\omega_{LO}-\omega_c)/k_BT)$ which agrees with the
exponential laws found for $N=2$ and $N=3$.

\section{Summary}

We have extended the theory for ODMPR to three-dimensional (3D) systems and
present the first detailed theoretical study of the magnetophonon resonance
in the frequency-dependent conductivity of electrons in bulk GaAs. In
comparison to the corresponding 2D systems, the theoretical obtained
amplitudes for the oscillations of both the linewidth and the effective mass
in a 3D system are for GaAs predicted to be about half of those in 2D.
Therefore, we believe that ODMPR can also be observed experimentally in bulk
polar semiconductors. Out numerical results indicate that the amplitude of
the resonant peak in the FWHM can be described by an exponential law 
for not too large temperatures.

\acknowledgments
This work was supported by FAPESP, CNPq (Brazil) and FWO, IUAP (Belgium).

\begin{figure}[tbp]
\caption{ (a) Re$\Sigma(\protect\omega)$ and (b) Im$\Sigma(\protect\omega)$
as a function of frequency $\protect\omega$ in GaAs at different magnetic
fields $\protect\omega_c/\protect\omega_{LO}=$0.3 (dotted curves), 0.4
(dashed curves) and 0.5 (solid curves). The corresponding absorption
spectrum is given in (c). The broadening parameter $\protect%
\gamma=0$ and temperature $T=77$ K.}
\label{fig1}
\end{figure}

\begin{figure}[tbp]
\caption{ The magneto-optical absorption spectrum around (a) $\protect%
\omega_c/\protect\omega_{LO}=1/2$ and (b) $\protect\omega_c/\protect\omega%
_{LO}=1/3$ for $T=77$ K and $\protect\gamma=0$.}
\label{fig2}
\end{figure}

\begin{figure}[tbp]
\caption{ The magneto-optical absorption spectra as a function of frequency $%
\protect\omega$ in GaAs for $\protect\gamma/\protect\omega_{LO}=0$ (solid
curve), 0.001 (dash curve), 0.01 (dotted curve) and 0.1 (dash-dotted
curve) at $\protect\omega_c/\protect\omega_{LO}=0.5$ and $T=77$ K.}
\label{fig3}
\end{figure}

\begin{figure}[tbp]
\caption{ (a) Polaron CR mass and (b) FWHM (full linewidth at half-maximum)
as a function of $\protect\omega_c$ at different temperatures from 60 K to
200 K with broadening $\protect\gamma/\protect\omega_{LO} =0.05$. }
\label{fig4}
\end{figure}

\begin{figure}[tbp]
\caption{ The CR mass oscillation amplitude as a function of temperature
at $\protect\omega _{c}/\protect\omega _{LO}=1/2$ (dots) and $\protect\omega %
_{c}/\protect\omega _{LO}=1/3$ (solid squares) with $\protect\gamma /\protect%
\omega _{LO}=0.05$.}
\label{fig5}
\end{figure}

\begin{figure}[tbp]
\caption{ An activation plot of the amplitude of the resonant peak in the
FWHM at $\protect\omega_c/\protect\omega_{LO}=1/2$ (circles) and $\protect%
\omega_c/\protect\omega_{LO}=1/3$ (triangles) as a function of $T^{-1}$ for
$\gamma/\omega_{LO}=0.05$. The
solid line $\propto \exp(-\hbar\protect\omega_{LO}/2k_BT)$ and the dotted
line $\propto \exp(-2\hbar\protect\omega_{LO}/3k_BT)$.}
\label{fig6}
\end{figure}


\bigskip 

\begin{references}
\bibitem{Barker}  J. R. Barker, J. Phys. C {\bf 5}, 1657 (1972).

\bibitem{RJN85}  R. J. Nicholas, Prog. Quantum Electron. {\bf 10}, 1 (1985);
R. V. Parfen'ev, G. I. Kharus, I. M. Tsidil'kovskii, and S. S. Shalyt, Sov.
Phys. -Usp. {\bf 17}, 1 (1974).

\bibitem{JVR77} J. Van Royen, J. De Sitter, L.F. Lemmens, and J.T. Devreese,
Physica B {\bf 89}, 101 (1977); J. Van Royen, J. De Sitter, and J.T. Devreese,
Phys. Rev. B {\bf 30}, 7154 (1984); J.P. Vigneron, R. Evrard, and E.
Kartheuser, Phys. Rev. B {\bf 18}, 6930 (1978).

\bibitem{DS}  D. Schneider, C. Brink, G. Irmer, and P. Verma, Physica B {\bf %
256-258}, 625 (1998); D. Schneider, K. Pricke, J. Schulz, G. Irmer, and M.
Wenzel, in {\it Proceedings of the 23rd International Conference on the
Physics of Semiconductors}, Eds. M. Scheffler and R. Zimmermann (World
Scientific, Singapore, 1996), p. 221.

\bibitem{PW88}  P. Warmenbol, F. M. Peeters, and J. T. Devreese, Solid State
Electron. {\bf 31}, 771 (1988); {\bf 32}, 1545 (1989); Phys. Rev. B {\bf 37}%
, 4694 (1988).

\bibitem{WX}  W. Xu, F. M. Peeters, J. T. Devreese, D. R. Leadley, and R. J.
Nicholas, Int. J. Mod. Phys. B {\bf 10}, 169 (1996); D. R. Leadley, R.J.
Nicholas, J. Singleton, W. Xu, F.M. Peeters, J.T. Devreese, J.A.A.J. Perenboom,
L. Van Bockstal, F. Herlach, J.J. Harris, and C.T. Foxon.
Phys. Rev. Lett. {\bf 73}, 589 (1994).

\bibitem{RJN90}  R. J. Nicholas, in {\it Landau Level Spectroscopy}, Eds. 
by G. Landwehr and E. I. Rashba (North-Holland, Amsterdam, 1990).

\bibitem{Wu97}  X.-G. Wu and F. M. Peeters, Phys. Rev. B 55, 9333 (1997);
{\it ibid.} {\bf 34}, 8800 (1986).

\bibitem{PV89}  P. Vasilopoulos, P. Warmenbol, F. M. Peeters, and J. T.
Devreese, Phys. Rev. B {\bf 40}, 1810 (1989).

\bibitem{GP}  G. Ploner, J. Smoliner, G. Strasser, M. Hauser, and Gornik,
Phys. Rev. B {\bf 57}, 3966 (1998).

\bibitem{DJB}  D. J. Barnes, R. J. Nicholas, F. M. Peeters, X. G. Wu, J. T.
Devreese, J. Singleton, C. J. G. M. Langerak, J. J. Harris, and Foxon, Phys.
Rev. Lett. {\bf 66}, 749 (1991).

\bibitem{FP83}  F. M. Peeters and J. T. Devreese, Phys. Rev. B {\bf 28},
6501 (1983).

\bibitem{FP86}  F. M. Peeters and J. T. Devreese, Phys. Rev. B {\bf 34},
7246 (1986).

\bibitem{HPD}  G. Q. Hai, F. M. Peeters, and J. T. Devreese, Phys. Rev. B 
{\bf 47}, 11358 (1993).
\end{references}
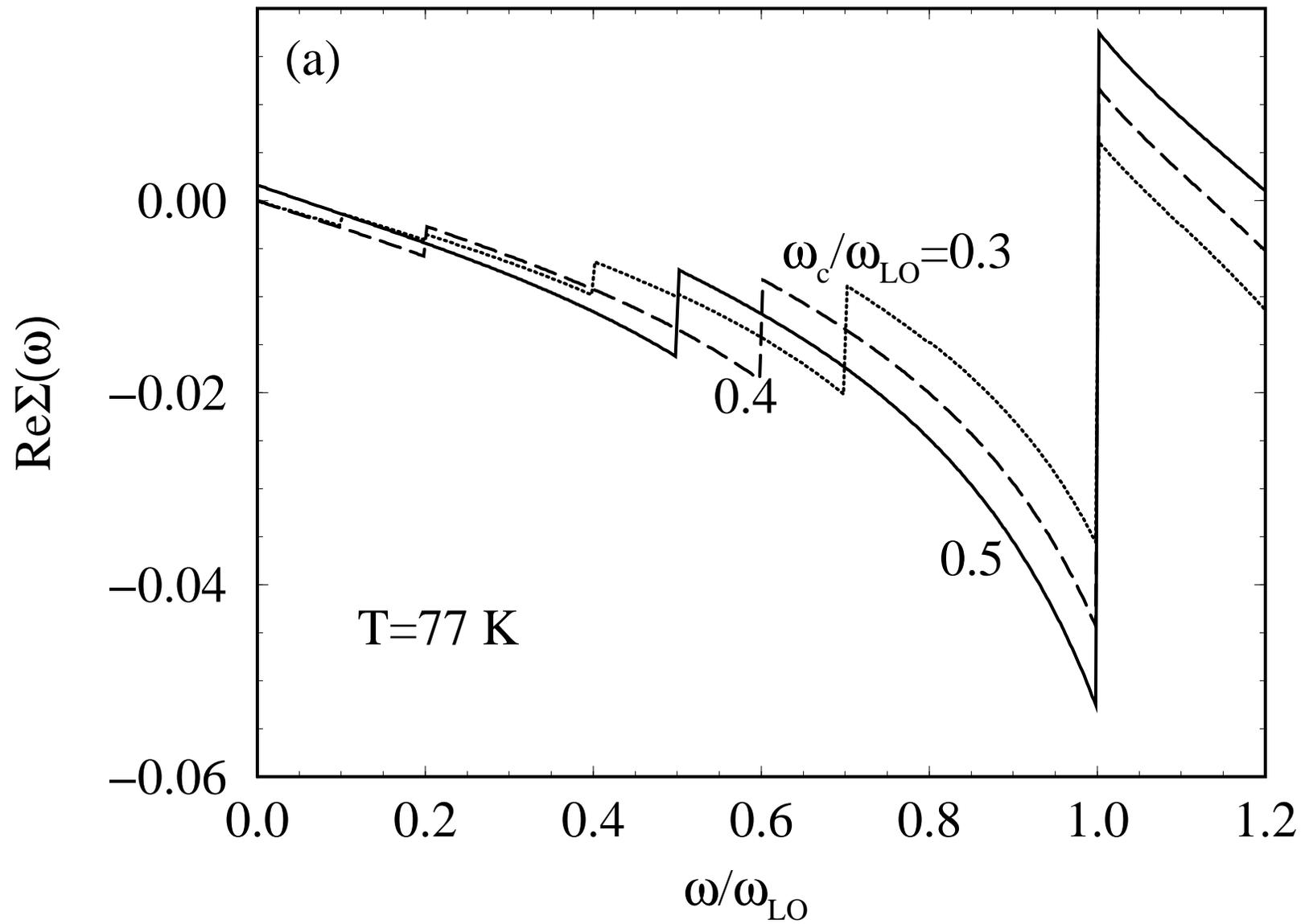
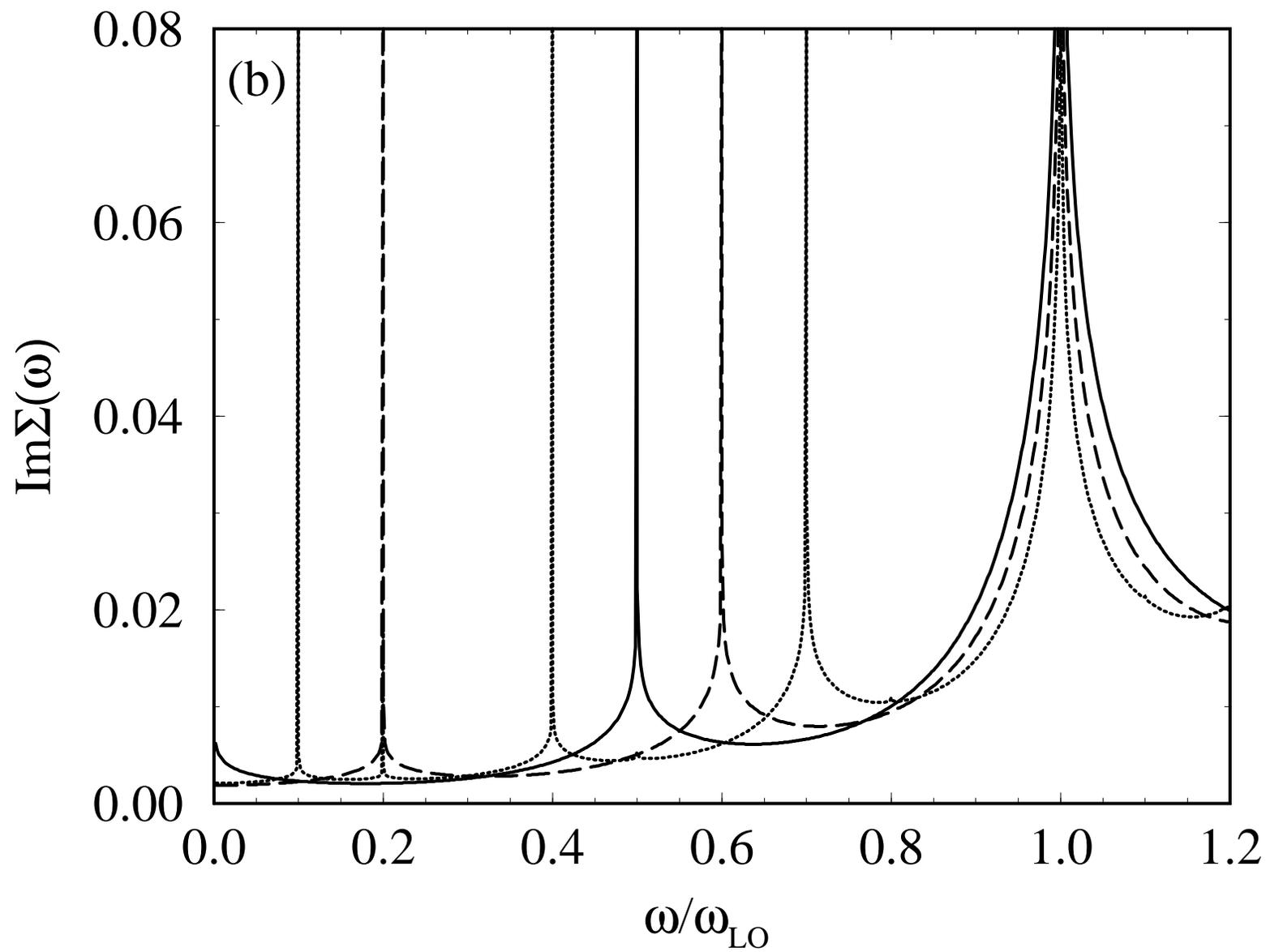
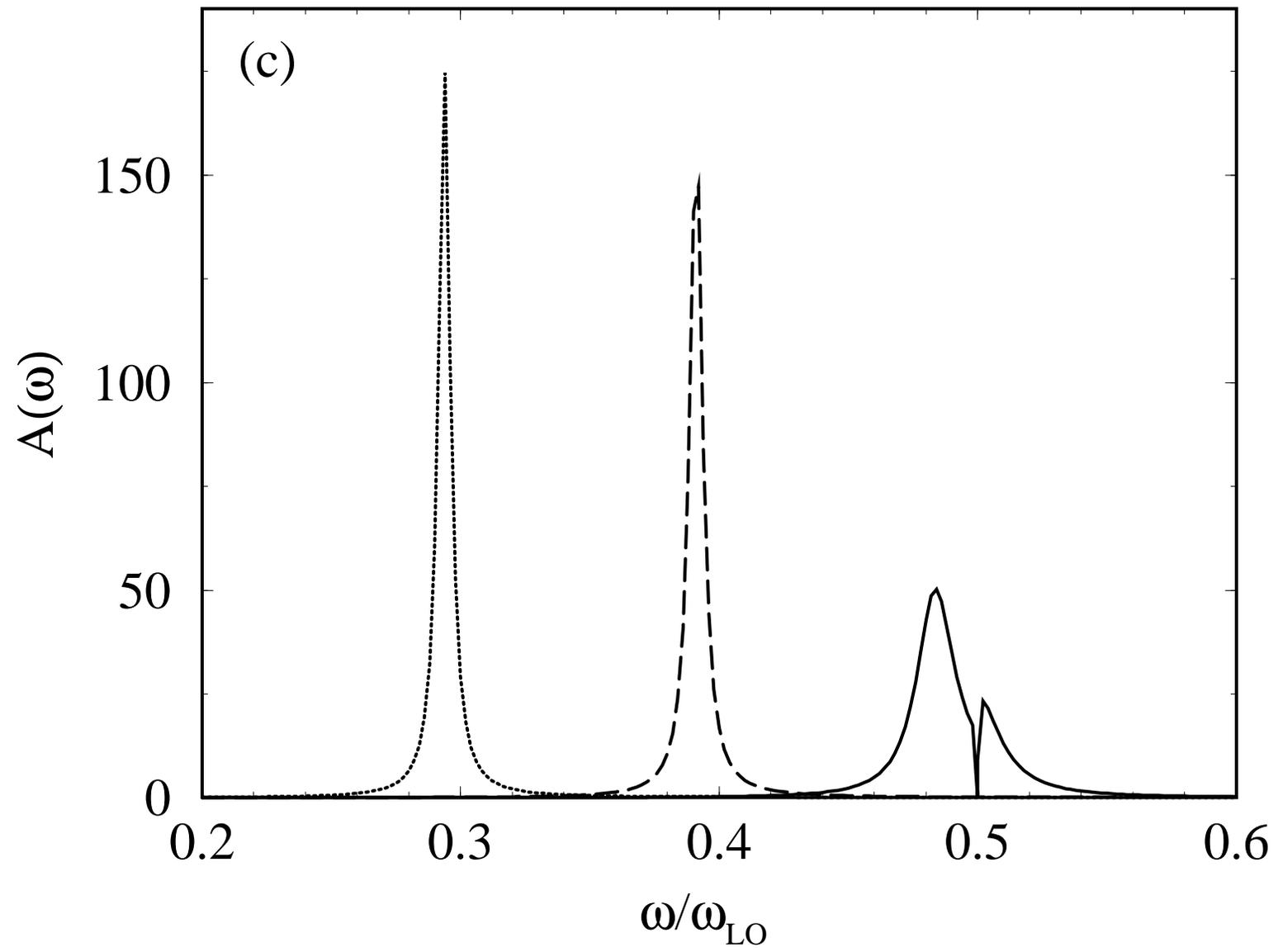
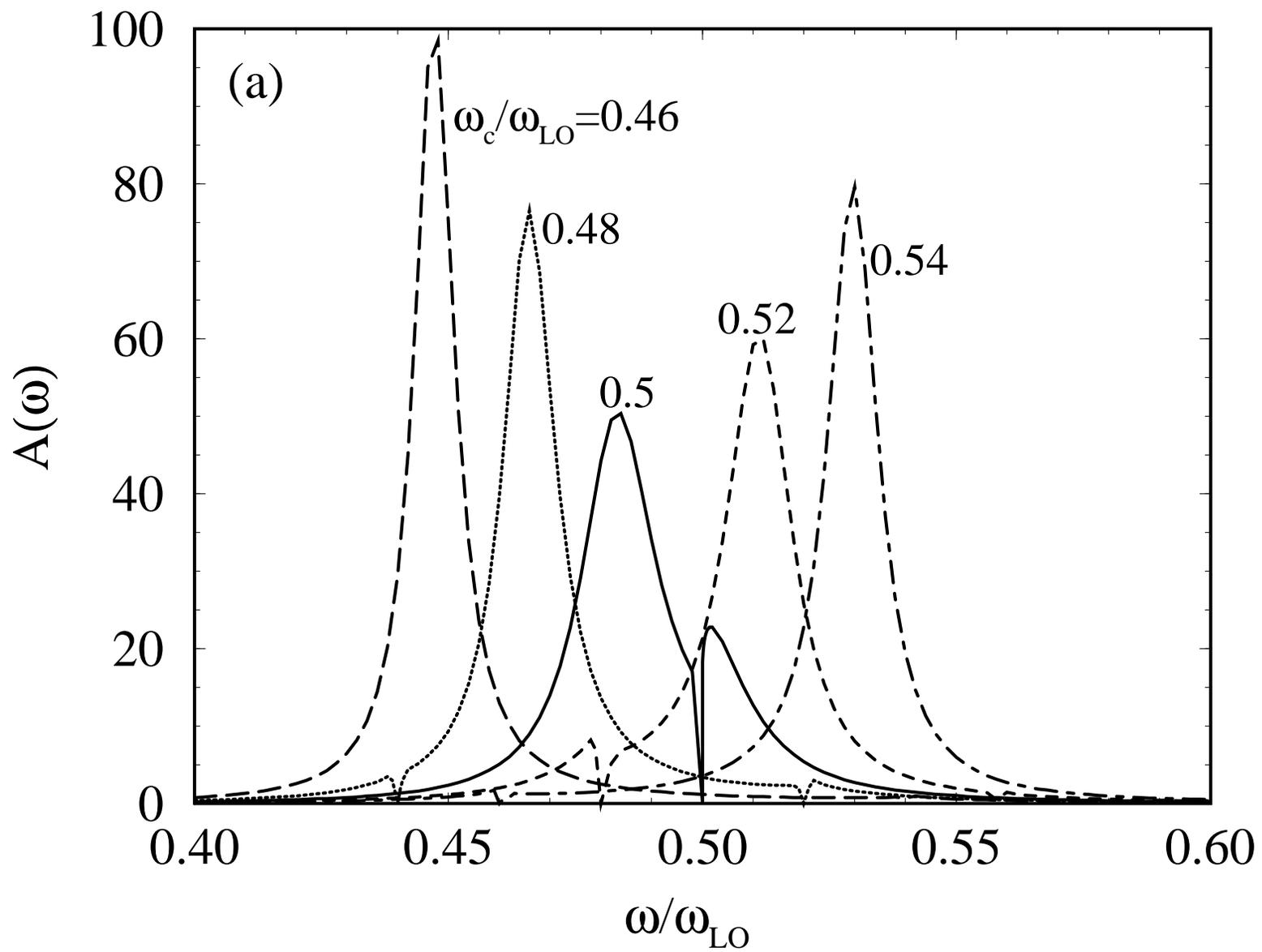
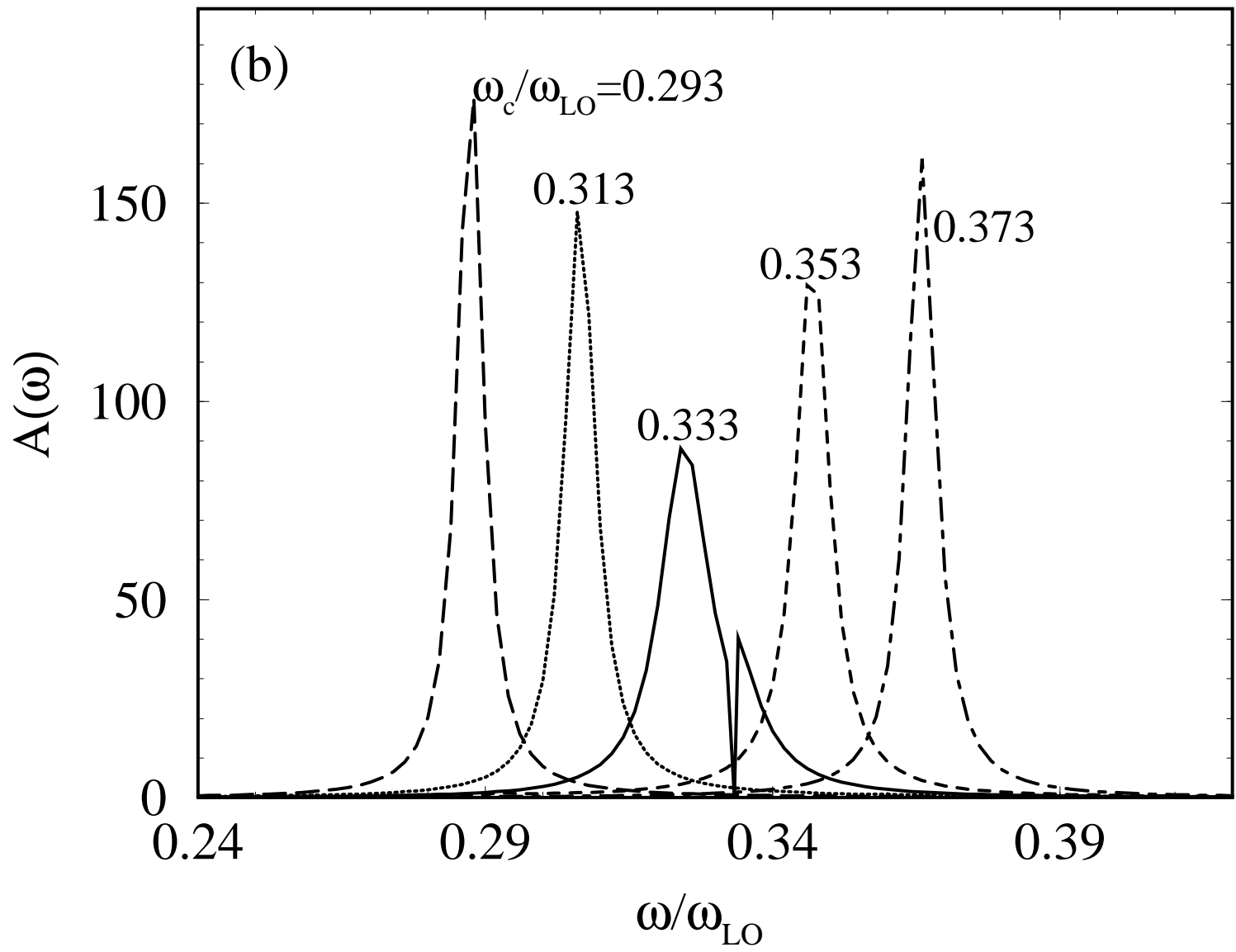
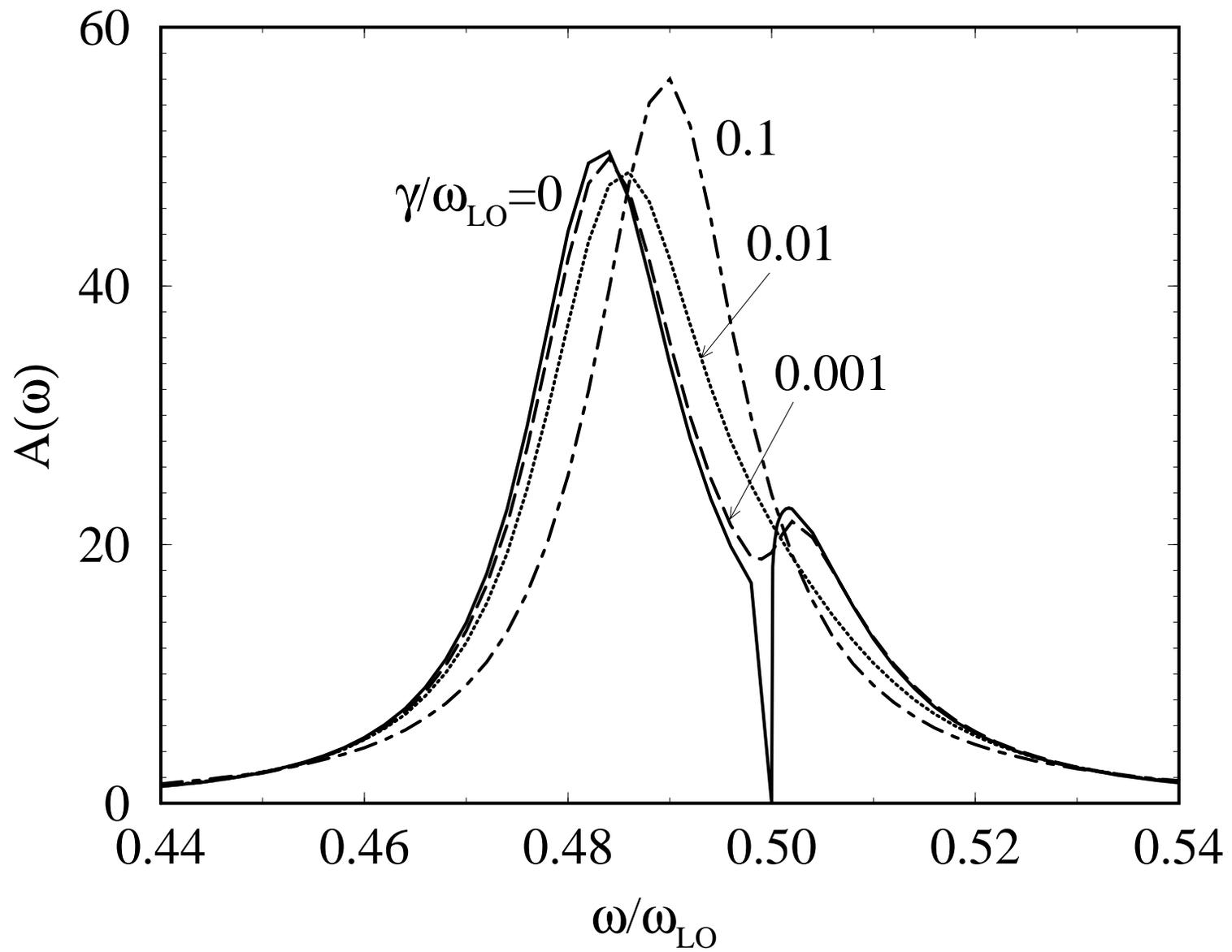
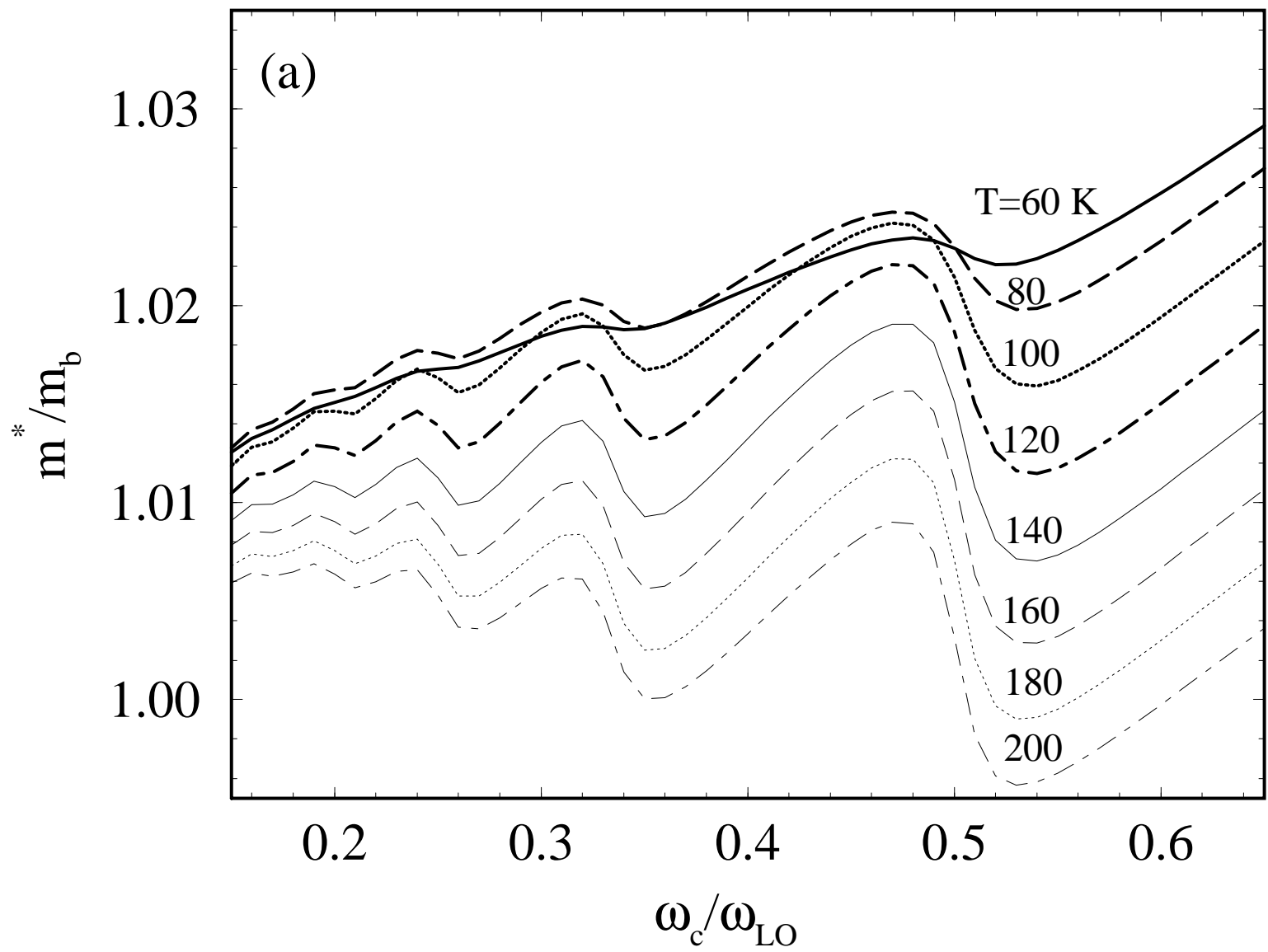
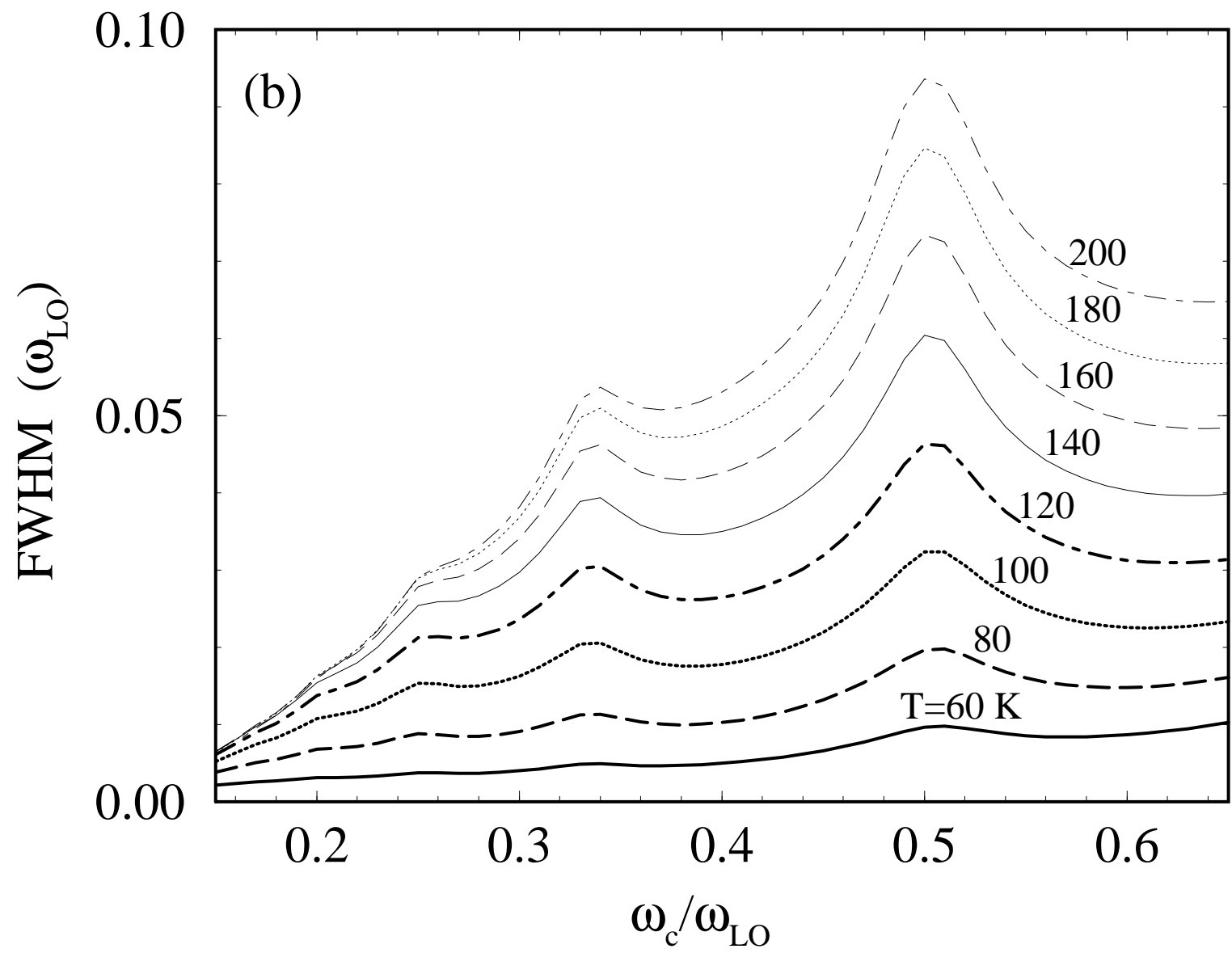
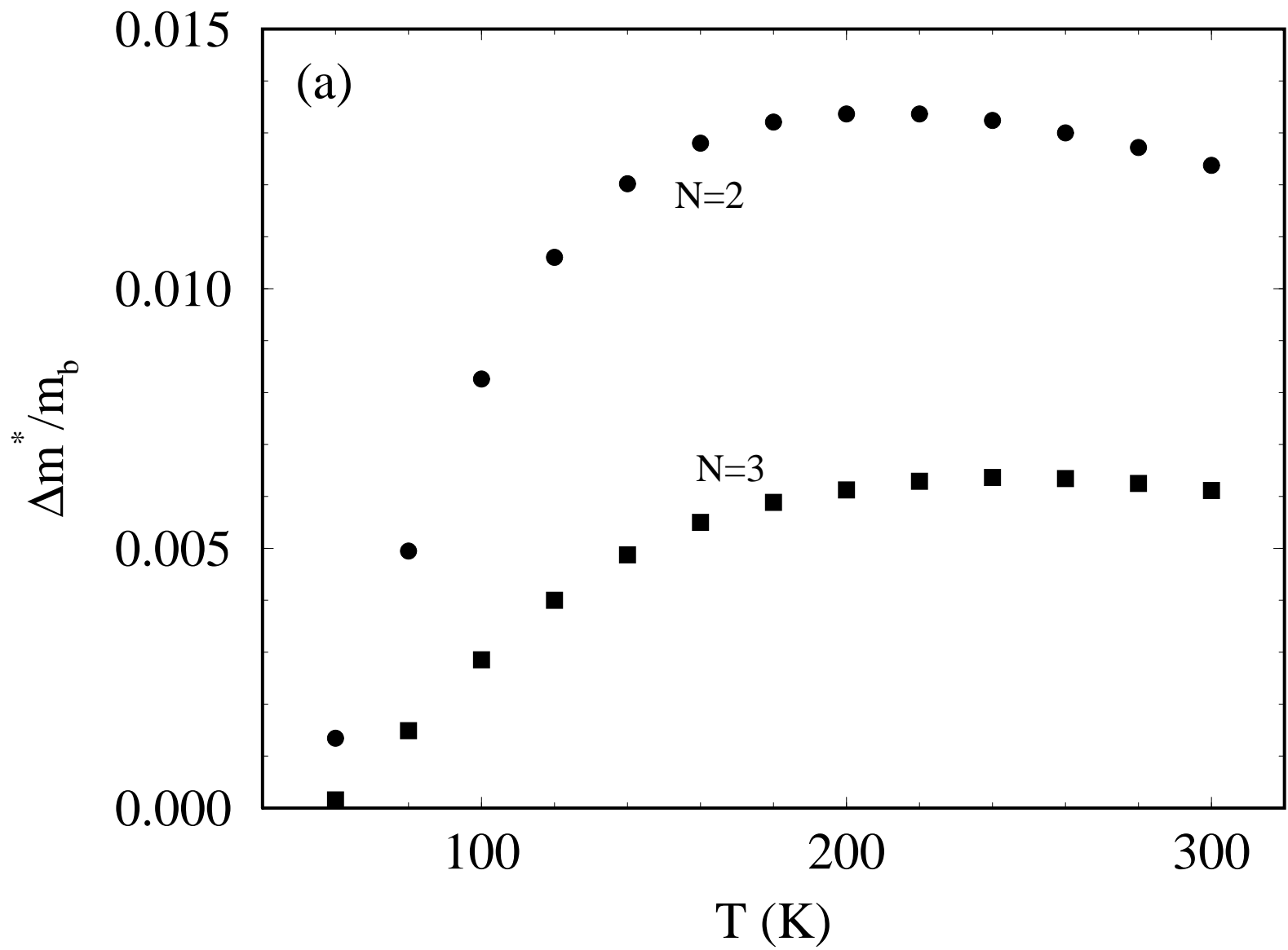
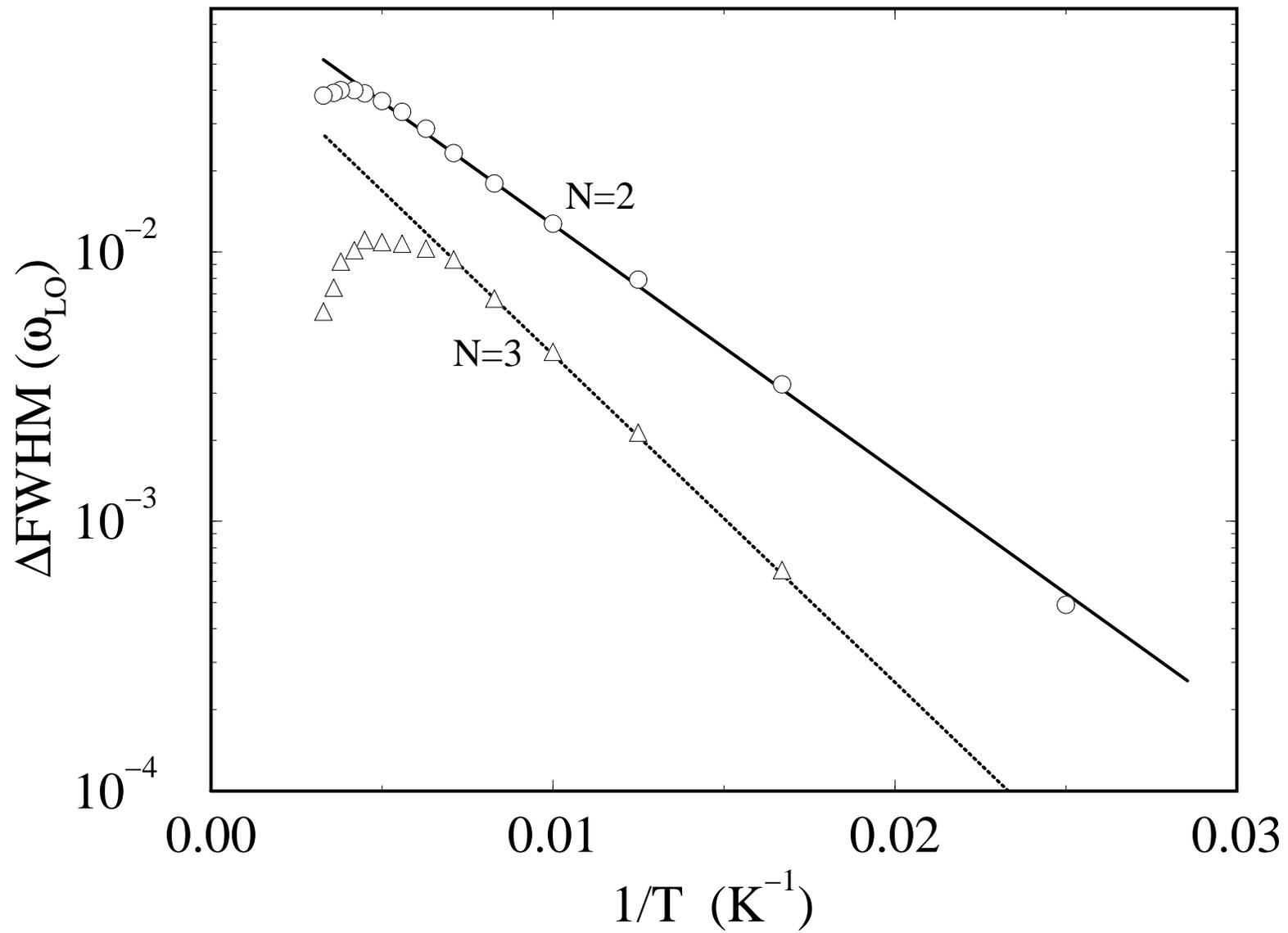
\end{document}